\documentclass[a4paper,11pt,oneside]{article}
\pdfoutput=1 
\usepackage{jheppub} 
\usepackage[backgroundcolor=white,textcolor=gray,linecolor=orange,bordercolor=orange,colorinlistoftodos,textsize=footnotesize]{todonotes}
\usepackage[T1]{fontenc} 
\usepackage{physics}
\usepackage{float}

\usepackage[latin9]{inputenc}

\usepackage[english]{babel}
\usepackage{fancyhdr}
\usepackage{braket}
\usepackage{caption}
\usepackage{subcaption}
\usepackage[compat=1.1.0]{tikz-feynman}

\title{One-loop algebras and fixed flow trajectories\\ in adjoint multi-scalar gauge theory}

\abstract{
We 
study the one loop renormalisation of 4d $SU(N)$ Yang-Mills theory with $M$ adjoint representation scalar multiplets related by $O(M)$ symmetry. General $M$ are of field theoretic interest, and the 4d one loop beta function of the gauge coupling $g^2$ vanishes for the case $M=22$, which is intriguing for string theory. This case is related to D3 branes of critical bosonic string theory in $D=22+4=26$. An RG fixed point could have provided a definition for a purely bosonic AdS/CFT, but we show that scalar self-couplings $\lambda$ ruin one-loop conformal invariance in the large $N$ limit. There are real fixed flows (fixed points of $\lambda/g^2$) only for $M\ge 406$, rendering one-loop fixed points of the gauge coupling and scalar couplings incompatible.

We develop and check an algebraic approach to the one-loop renormalisation group which we find to be characterised by a non-associative algebra of marginal couplings. In the large $N$ limit, the resulting RG flows typically suffer from strong coupling in both the ultraviolet and the infrared. Only for $M\ge 406$ fine-tuned solutions exist which are weakly coupled in the infrared. 
}

 \author{Nadia Flodgren}
 \author{and Bo Sundborg}
 \affiliation{The Oskar Klein Centre for Cosmoparticle Physics \& Department of Physics,\\
Stockholm University,\\
AlbaNova, 106 91 Stockholm, Sweden}

\emailAdd{nadia.flodgren@fysik.su.se}
\emailAdd{bo.sundborg@fysik.su.se}

\begin{document} 
\maketitle
\flushbottom

\section{Introduction}

Two classes of renormalisation group (RG) fixed points have been found amenable to perturbative treatment: infrared (IR) fixed points of the Caswell-Banks-Zaks type \cite{Caswell:1974gg,Banks:1981nn} and ultraviolet (UV) fixed points of the Litim-Sannino or asymptotically safe type \cite{Litim:2014uca,Litim:2015iea}. Both classes of fixed points involve gauge theory, and it seems that non-Abelian gauge theory is needed for weak coupling UV consistency \cite{Coleman:1973sx,Bond:2018oco}. In both cases, perturbative control is achieved through Veneziano type limits with a large number of colours $N$ and fermion flavours $N_F$. We study a similar, but purely bosonic theory, in order to see explicitly what happens in the absence of fermions.

To select scalar matter we appeal to simplicity: since a gauge theory is needed to improve short-distance behaviour, and the gauge bosons belong to the adjoint representation, it is simplest to consider scalars which also transform in the  
adjoint representation. We however allow for $M$ such scalar multiplets. Again, we choose the simplest versions of such theories, which are also symmetric under  global $O(M)$ transformations.

In addition to their simplicity, adjoint scalar gauge theories are similar to gauge theories derived from D3 branes of string theory, which have been studied before in the hope of finding non-supersymmetric conformal field theories. An early example is a non-supersymmetric D3-brane theory \cite{Klebanov:1999ch} based on type 0 string theory. For that theory, and other similar theories based on `orbifolding' $\mathcal{N}=4$ super-Yang-Mills theory \cite{Kachru:1998ys,Lawrence:1998ja,Bershadsky:1998cb} the gauge coupling and single trace beta functions vanish on a line of fixed points for large $N$, but it was soon realised that conformal invariance is ruined by double trace couplings, unless their values are complex \cite{Tseytlin:1999ii,Adams:2001jb,Dymarsky:2005uh,Pomoni:2009joh}. The present work will display both similarities and differences between our simple bosonic model and the superstring-related models.

Our present perspective is however complementary and almost opposite. We study gauge theories with purely scalar matter because they are in a sense the worst and most generic case for quantum properties of a gauge theory. 

Methodologically, we use specialised standard expressions \cite{Machacek:1984zw} and compute by hand, but we have also fixed normalisations so as to apply the Mathematica package RGBeta \cite{Thomsen:2021ncy} and compare results. We find complete agreement.

We also develop a new (to us) algebraic way of expressing the one loop RG equations. The simple basic observation is that the renormalisable four point couplings form a closed system determining their own RG flow. The one loop terms are second order and define a multiplication between linear combinations of couplings. The algebra defined this way turns out to be commutative and nonassociative, and completely encodes the one loop RG.
Although we hope that this algebraic picture can be extended to higher loops, we have not checked this, and do not make any such claims\footnote{Of course, for the case of a single scalar with a $\lambda \phi^4$ interaction, the algebra is associative, and higher loops are usefully expressed in powers of $\lambda$.}. Already the present one loop formulations opens new perspectives on multi-coupling RG equations and the structure of their  RG flows, which we explain in the body of the paper.

We search for fixed points of the RG flow for $M=22$, a case of special interest, since the gauge coupling beta function vanishes. This has been know for a long time, and is actually related to special properties of the bosonic string in $D=26 = 4+22$ dimensions \cite{Curtright:1981wv,Fradkin:1982kf,Nepomechie:1983cq,Dubovsky:2018vde}. It is even noted in \cite{Fradkin:1982kf} that the scalar coupling is renormalised. Now we can finally fill a gap in the analysis and determine non-trivial fixed points of the full set of couplings. None of the roots of the equations are real however, and all one-loop fixed point theories are therefore non-unitary. Instead requiring real fixed points, we find fixed points of relative couplings which represent `fixed flows' for the original scalar couplings, ie flows where all couplings run as the gauge coupling $g^2$. Such fixed flows however only appear for real couplings if $M\ge406$, corresponding to a large number of dimensions in a string interpretation.

In conclusion, we have developed an algebra description of the RG at one loop which could be interesting to extend to higher orders. We have also conclusively ruled out simultaneous real fixed points for all couplings in the multi-scalar models we explore. There are no theories in this class that are under one loop control in the UV, and only isolated cases that are under control in the IR. 

The paper is organised as follows. Section \ref{sect:quartic} reviews scalar gauge theories, their marginal interaction terms and standard expressions for their one loop beta functions. Then, we rephrase the general structure in terms of a commutative, but generically non-associative algebra. The algebraic reformulation reappears in other sections but is not necessary for appreciating the results. In section \ref{sect:msg}, the focus is  multi-adjoint scalar gauge theory and  the invariant polynomials or tensor structures that describe its interactions. A special subsection \ref{sect:algebra} is devoted to the resulting algebra, and also exemplifies the significance of various properties of the algebra. The expressions for the beta functions, and different checks of them, are then given in section \ref{sect:Adjointmsgbeta}. The actual RG equations, their qualitative structure and the characterisation of their fixed points and fixed flows are collected in section \ref{sect:RGFlow}. We finish with a short general discussion and an outlook in the concluding section \ref{sect:conclusion}.

\section{Quartic scalar interactions in $D=4$, beta functions and algebras}\label{sect:quartic}

We begin by reviewing scalar gauge theories, their marginally relevant interaction terms and beta functions. Then, we rephrase this structure algebraically. We introduce a \emph{marginal algebra}, which encodes the renormalisation group flow of classically scale invariant theories, and reflect its properties. 

\subsection{Quartic interactions and beta functions}

We focus on classically scale invariant scalar gauge theories in four dimensions. Hence, we only include kinetic terms, quartic interaction terms and minimal coupling to non-Abelian pure Yang-Mills theory. Alternatively, we can regard this set of theories as high energy limits of models which also include masses and cubic couplings which are dimensionful.

The interaction terms in the Lagrangian are quartic polynomials, which are automatically symmetrised when written in terms of general reducible representations. 
We employ symbolic multi-indices ${\bar{A}\bar{B}\bar{C}\bar{D}}$ etcetera, which include indices labelling the individual irreducible representations as well as the corresponding gauge indices. Alternatively, we may view these indices as the indices of a single, typically reducible, representation of the gauge group. 

With this understanding, the total Lagrangian including Yang-Mills term in terms of the vector potential $A$, minimally coupled scalars and the explicit form of the quartic interaction term
reads
\begin{equation} \label{eq:lagrange}
\mathcal{L} = \mathcal{L}_{YM}[A] +\mathcal{L}_{0}[\phi,A]-\frac{1}{4!}\lambda_{\bar{A}\bar{B}\bar{C}\bar{D}}\phi_{\bar{A}}\phi_{\bar{B}}\phi_{\bar{C}}\phi_{\bar{D}}\ .
\end{equation}
The quartic interaction tensor $\lambda_{\bar{A}\bar{B}\bar{C}\bar{D}}$ is taken to be totally symmetric in the multi-indices. When constructing the $\lambda_{\bar{A}\bar{B}\bar{C}\bar{D}}$ tensor from specific invariant polynomials below we get specific symmetry factors for each invariant.

The one-loop beta functions of these theories are special cases of standard results collected for example in \cite{Machacek:1984zw,Luo:2002ti}:
\begin{equation} \label{eq:beta}
\begin{split}
&\frac{d}{dt}{\lambda}_{\bar{A}\bar{B}\bar{C}\bar{D}}=\mu \frac{d}{d\mu}\lambda_{\bar{A}\bar{B}\bar{C}\bar{D}}=\beta_{\bar{A}\bar{B}\bar{C}\bar{D}} =\frac{1}{(4\pi)^2}  \beta^I_{\bar{A}\bar{B}\bar{C}\bar{D}}\\
&\beta^I_{\bar{A}\bar{B}\bar{C}\bar{D}}= \Lambda_{\bar{A}\bar{B}\bar{C}\bar{D}}^2-3g^2\Lambda^S_{\bar{A}\bar{B}\bar{C}\bar{D}} +3g^4A_{\bar{A}\bar{B}\bar{C}\bar{D}}
\end{split}
\end{equation}
Here, $\mu$ is the scale, and we think of $t=\log \mu$ as the ``time'' in the RG equations, running from the IR to the UV.
The different terms arise from different classes of Feynman diagrams, with different group theory factors
\begin{equation} \label{eq_LLA}
\begin{split}
&\Lambda^2_{\bar{A}\bar{B}\bar{C}\bar{D}}=\frac{1}{8}\sum_{\text{perms}} \lambda_{\bar{A}\bar{B}\bar{E}\bar{F}}\lambda_{\bar{E}\bar{F}\bar{C}\bar{D}} \\
&\Lambda^S_{\bar{A}\bar{B}\bar{C}\bar{D}}=\sum_i C(i)\lambda_{\bar{A}\bar{B}\bar{C}\bar{D}} \\
&A_{\bar{A}\bar{B}\bar{C}\bar{D}} = \frac{1}{8}\sum_{\text{perms}} \{ \theta^E,\theta^F\}_{\bar{A}\bar{B}}\{ \theta^E,\theta^F\}_{\bar{C}\bar{D}},
\end{split}
\end{equation}
where the sum over $i$ is a sum over four external legs, $C(i)$ being the corresponding quadratic Casimir operator. Similarly, the sum over permutations is also a sum over permutations over the four external legs, $\theta^C_{\bar{A}\bar{B}}$ being a possibly reducible representation of the gauge Lie algebra and indices $A, B, C, \dots$ are Lie algebra indices. Clearly it is important in the use of the group theory expressions to use the same normalisation as in \cite{Machacek:1984zw}. We have concluded that the normalisation of matrices $T_A$ in the fundamental representation should be 
\begin{equation}
\operatorname{Tr} T_A T_B = T_R \delta_{AB}=\frac{1}{2} \delta_{AB} \ .
\end{equation}
From this normalisation all values of Casimir operators follow, eg the quadratic Casimir in the adjoint representation
\begin{equation}
C_2(G)=N\ .
\end{equation}
for the $G=SU(N)$ gauge group.

\subsection{Couplings and algebras of couplings}

We now introduce the \emph{marginal algebra} of marginal four point couplings to describe the one loop renormalisation of the theories of interest to us. We have found it an illuminating way to do the calculations and describe the structures in the theory, but the algebra is not essential to understand our results and we will also describe them in more conventional terms below.
 
The idea behind the algebra originates in renormalisation theory. In a renormalisable theory there is a finite number of couplings $\lambda$ that are renormalised. The renormalisation group (RG) couples their evolution under a change of scale $\mu$ in non-linear renormalisation group equations. Importantly, the RG equations are determined by beta functions $\beta(\lambda)$  
which are given by specific sums of Feynman diagrams.
Each diagram in this perturbation theory contributes with a monomial in the marginal couplings and the total RG equation defines flows $\lambda(t)$ in the space of couplings. 

Ignoring higher orders, since we are focusing on the one loop approximation, and thinking about the marginal couplings $\lambda$ as a vector space $\mathcal{A}$, the RG equations give a flow in this vector space, determined by a second order expression from the Feynman diagrams, symbolically 
\begin{equation}\label{eq:P_2}
\beta_{\lambda}(\lambda)=\frac{1}{(4\pi)^2}\left(P_2(\lambda) + P_1(\lambda)+ P_0\right)\\ ,
\end{equation}
where the subscript $q$ in $P_q$ denotes the order in $\lambda$. Note that the polynomials $P_q$ take values in the vector space of couplings. This structure naturally gives rise to an algebra with a product constructed  as
\begin{equation}\label{eq:def_diamond}
\lambda \diamond \kappa \equiv \frac{1}{2}\left(P_2(\lambda+\kappa)-P_2(\lambda)-P_2(\kappa)\right)\ ,
\end {equation}
to isolate the cross-terms and become linear in $\lambda$ and $\kappa$ separately. Recall that an algebra means a vector space equipped with a product satisfying the distributive law. From now on, we can therefore regard the space $\mathcal{A}$ of couplings as an algebra. By its definition, we see that the product is commutative, but note that there is no guarantee for it to be associative.

Let us now spell out how the algebra picture works concretely with the standard RG equations cited above.

Suppose that we have a complete basis $g_{\bar{A}\bar{B}\bar{C}\bar{D}}^k$ of symmetric tensors with corresponding coupling constants $\lambda_k$ such that the general symmetric couplings \eqref{eq:lagrange} and their beta functions \eqref{eq:beta} are captured by
\begin{equation} \label{eq_assume}
\begin{split}
\lambda_{\bar{A}\bar{B}\bar{C}\bar{D}} = \lambda_k g^k_{\bar{A}\bar{B}\bar{C}\bar{D}}
\end{split}
\end{equation}
\begin{equation} \label{eq_antz}
\begin{split}
\beta_{\bar{A}\bar{B}\bar{C}\bar{D}} = \beta_k g^k_{\bar{A}\bar{B}\bar{C}\bar{D}}.
\end{split}
\end{equation}
The beta functions at one-loop order
\begin{equation} 
\begin{split}
\beta_{\bar{A}\bar{B}\bar{C}\bar{D}}&= \beta_k g^k_{\bar{A}\bar{B}\bar{C}\bar{D}} \\
&=\frac{1}{(4\pi)^2}  \beta^I_{\bar{A}\bar{B}\bar{C}\bar{D}} =  \frac{1}{(4\pi)^2}(\Lambda_{\bar{A}\bar{B}\bar{C}\bar{D}}^2-3g^2\Lambda^S_{\bar{A}\bar{B}\bar{C}\bar{D}} +3g^4A_{\bar{A}\bar{B}\bar{C}\bar{D}}) \\
\end{split}
\end{equation}
can then be written in terms of the $g_{\bar{A}\bar{B}\bar{C}\bar{D}}^k$ basis as
\begin{equation} \label{eq_bk1}
\begin{split}
(4\pi)^2 \beta_k = ( \lambda_m\lambda_n  c^{mn}_k - 3g^2C_1 \lambda_k  + 3g^4d_k) ,\\
\end{split}
\end{equation}
where $C_1=4C_2(G)$ for four point vertices in the adjoint representation.The right hand side has been expanded in the complete basis. The coefficients $c^{mn}_k$ in the quadratic first term capture the whole quadratic contribution because the basis is assumed to be closed under the renormalisation group flow, the linear second term is a real multiple of the tree-level coupling because the wave-function renormalisation is diagonal in this model, and $d_k$ represents the RG flow of scalar couplings induced by other interactions, here the gauge coupling. The $c^{mn}_k$ act as structure constants which encode how a second order expression in the coupling constants $\lambda_m$ and $\lambda_n$ enter in the beta function for $\lambda_k$. This is the origin of the algebra description.

The algebra product
\begin{equation} \label{eq_diamond}
\begin{split}
\left(\lambda \diamond \kappa\right)_{\bar{A}\bar{B}\bar{C}\bar{D}} \equiv \frac{1}{8}\sum_{\text{perms}} \lambda_{\bar{A}\bar{B}\bar{E}\bar{F}}\ \kappa_{\bar{E}\bar{F}\bar{C}\bar{D}}  ,
\end{split}
\end{equation}
may be taken as the practical definition, which replaces \eqref{eq:def_diamond} by realising that equations \eqref{eq:P_2}, \eqref{eq:beta} and \eqref{eq_LLA} effectively equates $P_2$ with $\Lambda^2$.
The sum over permutations refers to all permutations of the four external legs, which guarantees that the product is totally symmetric, just as its two factors $\lambda$ and $\kappa$. For the basis couplings
\begin{equation} \label{eq_diamond}
\begin{split}
\left(g^m \diamond g^n\right)_{\bar{A}\bar{B}\bar{C}\bar{D}} = \frac{1}{8}\sum_{\text{perms}} g^m_{\bar{A}\bar{B}\bar{E}\bar{F}}g^n_{\bar{E}\bar{F}\bar{C}\bar{D}}  
  = c^{mn}_kg^k_{\bar{A}\bar{B}\bar{C}\bar{D}}
\end{split}
\end{equation}
which may be written abstractly as a product $g^m \diamond g^n  = c^{mn}_k g^k$ by suppressing all field indices. Since coupling tensors $g_{\bar{A}\bar{B}\bar{C}\bar{D}}^k$ form a linear space and the product is distributive the component version is manifestly a representation of an abstract algebra $\mathcal{A}$. In this representation the $\diamond$ product is commutative by the total symmetry and the sum over permutations. The coefficients $c^{mn}_k$ completely characterise the product. By construction the basis spans the space of symmetric 4-point couplings that is engaged by the renormalisation group flow, generated by the beta functions
\begin{equation} \label{eq_bk}
\begin{split}
(4\pi)^2 \beta_k = ( \lambda_m g^m \diamond \lambda_n  g^n)_k - 3g^2C_1 \lambda_k  + 3g^4d_k). \\
\end{split}
\end{equation}

We have established a form for one-loop RG equations for multiple couplings. It will be helpful to have a general understanding of transformation properties under general coordinate transformations in the space of couplings. We only have to remember that the beta functions are not scalars, but vectors in the space of couplings, by their definition \eqref{eq:beta}. Hence, the beta functions transform as
\begin{equation}
\beta_i (\lambda)= \mu \frac{d\lambda_i}{d\mu}(\lambda)=\mu \frac{\partial \lambda_i}{\partial \kappa_j}\frac{d\kappa_j}{d\mu}(\kappa)=\frac{\partial \lambda_i}{\partial \kappa_j} \beta_j(\kappa)\ .
\end{equation}
The coefficients in the second order expression \eqref{eq_bk} will then also transform as tensors in the space of coupling constants. The transformations
\begin{equation} \label{eq:trans}
\begin{split}
&\lambda_k \to \kappa_k=R_k^{{\ }m} \lambda_m, \lambda_m = R^k_{{\ }m}\kappa_k\\
&R_k^{{\ }m}R^k_{{\ }n}= \delta^m_n= R_m^{{\ }k}R^n_{{\ }k}\\ 
\end{split}
\end{equation}
yield
\begin{equation} \label{eq:transcoeff}
\begin{split}
&d_j \to d_j (\kappa)=d_i \frac{\partial \kappa_j}{\partial \lambda_i}=d_i R_j^{{\ }i} \\
&\lambda_m\lambda_n  c^{mn}_k \to \lambda_m\lambda_n  c^{mn}_i \frac{\partial \kappa_k}{\partial \lambda_i} = \lambda_m\lambda_n  c^{mn}_i R_k^{{\ }i}= \kappa_m \kappa_n R^m_{{\ }i} R^n_{{\ }j} c^{ij}_l R_k^{{\ }l} \\
& c^{mn}_k \to c^{mn}_k(\kappa)= R^m_{{\ }i} R^n_{{\ }j} c^{ij}_l R_k^{{\ }l} .\\
\end{split}
\end{equation}

\section{Adjoint multi-scalar gauge theory invariants and their algebra}\label{sect:msg}

We introduce adjoint multi-scalar gauge theory and represent its interactions in two different but related ways, as invariant polynomials and as symmetric tensor structures. We then discuss the resulting algebra $\mathcal{A}$ of marginal couplings before and after the large $N$ limit. Finally, we detail algebraic properties of $\mathcal{A}$ with physical interpretations for some of them. Specifically, we note some simplifications in the large $N$ limit.

\subsection{Adjoint multi-scalar gauge theory}

For multi-scalar gauge theories with all scalars in the adjoint representation of the gauge group, \emph{all} fields belong to the adjoint representation. A simple description is then obtained by writing all fields, including the vector potentials, as $N \times N$ matrices and constructing gauge invariants from traces of products of matrices. We choose to study $M$ adjoint scalar multiplets of an $SU(N)$ gauge group. The adjoint representation is real, permitting an $O(M)$ symmetry to act on the scalar multiplets. In effect, the index $a$ labelling the scalar $SU(N)$ matrix fields $\Phi_a$ transforms in the fundamental representation of $O(M)$.

The Lagrangian of the theory may be summarised as
\begin{equation}
\mathcal{L}= - \frac{1}{4}\operatorname{Tr} F^{\mu\nu}F_{\mu\nu} + \frac{1}{2} \operatorname{Tr} D^\mu \Phi_a D_\mu \Phi_a - \frac{1}{4!}  \operatorname{Tr} P_4(\Phi)\ ,
\end{equation}
where $F$ is the Yang-Mills field strength and $D$ is the gauge covariant derivative. $P_4$ is a homogeneous fourth order polynomial in $M$ variables, which we will take to be $O(M)$ invariant. It is in general a sum over four independent invariants which we describe below.
The different ways to write the scalars are related by
\begin{equation}
\Phi_a = \phi_{aA}T_A = \phi_{\bar{A}}T_A \ ,
\end{equation}
which also explains how we resolve the multi-indices $\bar{A}$.

Then the unique quadratic invariant
\begin{equation}\label{eq:quadratic}
\mathcal{O}_2 = \operatorname{Tr} \Phi_a \Phi_a = \delta_{ab} \operatorname{Tr} \phi_{aA}T_A  \phi_{bB}T_B  = \frac{1}{2}\delta_{ab} \delta_{AB}  \phi_{aA} \phi_{bB}
\end{equation}
illustrates what will be a recurring theme below. Invariants may be represented either as invariant polynomials, $\operatorname{Tr} \Phi_a \Phi_a$, or as invariant tensor structures, $\frac{1}{2}\delta_{ab} \delta_{AB} $.

\subsection{Invariant polynomials}
We now turn to the invariants needed to describe an $O(M)$ and $SU(N)$ symmetric potential. By squaring $\mathcal{O}_2$ and ordering it to remove short-distance singularities we can define a quartic double trace operator $\operatorname{Tr} \Phi_a \Phi_a \operatorname{Tr} \Phi_b \Phi_b$. We will first discuss more general renormalisable terms in the potential written as quartic \emph{polynomials} of the fields, and then in terms of the $SU(N)$ and $O(M)$ \emph{tensor structures} which summarise the coefficients of the polynomials.

The list of invariant quartic polynomials includes $\operatorname{Tr} \Phi_a \Phi_a \operatorname{Tr} \Phi_b \Phi_b$ and $\operatorname{Tr} \Phi_a \Phi_b \operatorname{Tr} \Phi_a \Phi_b$, which are both double trace expressions - sums of squares of $SU(N)$ traces. There are also two different single trace monomials. The full space of invariants is spanned by the operators 
\begin{equation}\label{eq:inv_pol}
\operatorname{Tr} \Phi_a \Phi_a \Phi_b \Phi_b ,
\operatorname{Tr} \Phi_a \Phi_b \Phi_a \Phi_b ,
\operatorname{Tr} \Phi_a \Phi_a \operatorname{Tr} \Phi_b \Phi_b ,
\operatorname{Tr} \Phi_a \Phi_b \operatorname{Tr} \Phi_a \Phi_b \ .
\end{equation} 
The double trace polynomials are important for several reasons, primarily because they decompose, as above, into factors of quadratic polynomial invariants of the gauge group $SU(N)$. The large $N$ scaling of single and double trace operators is quite different, and leads to important differences in the large $N$ limit \cite{Pomoni:2009joh}. Double trace operators appear as corrections to single trace potentials, but are still significant for large $N$. We shall see this in in more detail below. In particular, we will find that the algebra of marginal operators differentiates clearly between single trace and double trace operators.

\subsection{Invariant tensor structures}

To each invariant polynomial corresponds an invariant tensor structure, which we label by the same symbols. It is obtained by expressing the $\Phi_a$ in terms of the Lie algebra basis and identifying the coefficients in the quartic polynomial of $\phi_{aA}$. Conversely, we may start from the completely symmetric tensor structures  
\begin{equation} \label{eq_base}
\begin{split}
{g}^{1s} &\equiv \delta_{ab}\delta_{cd}( \Tr(T_AT_BT_CT_D) + \Tr(T_AT_DT_CT_B) +\Tr(T_AT_BT_DT_C) + \Tr(T_AT_CT_DT_B)  ) \\
&+ \delta_{ac}\delta_{bd}(\Tr(T_AT_BT_DT_C) + \Tr(T_AT_CT_DT_B) + \Tr(T_AT_CT_BT_D) +\Tr( T_AT_DT_BT_C)) \\
&+ \delta_{ad}\delta_{bc}( \Tr(T_AT_CT_BT_D) + \Tr(T_AT_DT_BT_C) +  \Tr(T_AT_BT_CT_D) + \Tr(T_AT_DT_CT_B)) \\
{g}^{1t} &\equiv \delta_{ab}\delta_{cd}(\Tr(T_AT_CT_BT_D) + \Tr(T_AT_DT_BT_C)) + \delta_{ac}\delta_{bd}(\Tr(T_AT_BT_CT_D) + \Tr(T_AT_DT_CT_B)) \\
&+ \delta_{ad}\delta_{bc}(\Tr(T_AT_BT_DT_C) + \Tr(T_AT_CT_DT_B)) \\ 
{g}^{2s}&\equiv\delta_{a b} \delta_{c d} \delta_{A B} \delta_{C D}+\delta_{a c} \delta_{b d} \delta_{A C} \delta_{B D}+\delta_{a d} \delta_{b c} \delta_{A D} \delta_{B C} \\
{g}^{2t}&\equiv \delta_{a b} \delta_{c d}(\delta_{A C} \delta_{B D}+\delta_{A D} \delta_{B C})
+ \delta_{a c} \delta_{b d}(\delta_{A B} \delta_{C D}+\delta_{A D} \delta_{C B})
+ \delta_{a d} \delta_{b c}(\delta_{A B} \delta_{D C}+\delta_{A C} \delta_{D B}), \\
\end{split}
\end{equation}
where we have suppressed field indices. The polynomials follow from contracting the tensor structures above with the scalar fields $\phi_{aA}$. The superscript 1 and 2 denote single trace and double trace objects, respectively, while we use $s$ to symbolise scalar product and $t$ to symbolise tensor products of neighbouring $O(M)$ vectors. The notation may become clearer in the expressions for the corresponding invariant polynomials

\begin{equation}
\begin{aligned}\label{eq:inv_pol}
\frac{1}{4!}g^{1 s}(\Phi)\equiv \frac{1}{4!} {g}^{1s}_{\bar{A}\bar{B}\bar{C}\bar{D}}\phi_{\bar{A}} \phi_{\bar{B}}\phi_{\bar{C}}\phi_{\bar{D}}  &=\frac{1}{2}\operatorname{Tr} \Phi_a \Phi_a \Phi_b \Phi_b 
\\
\frac{1}{4!}g^{1 t}(\Phi)\equiv \frac{1}{4!} {g}^{1t}_{\bar{A}\bar{B}\bar{C}\bar{D}}\phi_{\bar{A}} \phi_{\bar{B}}\phi_{\bar{C}}\phi_{\bar{D}}  &=\frac{1}{4}\operatorname{Tr} \Phi_a \Phi_b \Phi_a \Phi_b 
\\
\frac{1}{4!}g^{2 s}(\Phi) \equiv \frac{1}{4!} {g}^{2s}_{\bar{A}\bar{B}\bar{C}\bar{D}}\phi_{\bar{A}} \phi_{\bar{B}}\phi_{\bar{C}}\phi_{\bar{D}} &=\frac{1}{8}\operatorname{Tr} \Phi_a \Phi_a \operatorname{Tr} \Phi_b \Phi_b 
\\
\frac{1}{4!}g^{2 t}(\Phi) \equiv \frac{1}{4!} {g}^{2t}_{\bar{A}\bar{B}\bar{C}\bar{D}}\phi_{\bar{A}} \phi_{\bar{B}}\phi_{\bar{C}}\phi_{\bar{D}} &=\frac{1}{4}\operatorname{Tr} \Phi_a \Phi_b \operatorname{Tr} \Phi_a \Phi_b \ ,\\
\end{aligned}
\end{equation} 
which comply with the form \eqref{eq:lagrange}, and the factors in front of the trace expressions are  symmetry factors obtained from the number of terms in the totally symmetric tensors \eqref{eq_base}. Any invariant tensor structure $\lambda_{\bar{A}\bar{B}\bar{C}\bar{D}}$ with total symmetry may then be expanded in the basis of invariant tensor structures:
\begin{equation}\label{eq:decomposition}
\lambda_{\bar{A}\bar{B}\bar{C}\bar{D}}=\lambda_k g^k_{\bar{A}\bar{B}\bar{C}\bar{D}}\ .
\end{equation}
The most important factor above is actually the $1/4!$ which establishes the connection to standard beta function results \eqref{eq:beta}. Then, any change of tensor structure basis will translate to a different algebra and different coupling constants and beta functions, but always respecting the equations in section \ref{sect:quartic}.

\subsection{Algebra}\label{sect:algebra}

The commutative but non-associative algebra $\mathcal{A}$ defined above in equation \eqref{eq_diamond} can be computed by any method for multiplying tensors, contracting their indices and identifying the resulting objects\footnote{In practice, we used a graphical method, inspired by birdtrack calculus, \cite{Cvitanovic:1976am,Cvitanovic:2008zz}
. 
Our results have been checked and compared by RGBeta which is Mathematica packge, \cite{Thomsen:2021ncy}.}.

The single trace tensor multiplication table reads
\begin{equation}\label{eq:1trdiamond}
\begin{split}
g^{1s} \diamond g^{1s} &= \frac{1}{2}(M+3)g^{2s}+\frac{1}{2}g^{2t}+\frac{1}{2}N(M+3)g^{1s} \\ 
g^{1t} \diamond g^{1t} &= \frac{1}{8}(M+2)g^{2t}+\frac{1}{2}Ng^{1s} \\
g^{1s} \diamond g^{1t} &= \frac{1}{2}g^{2s}+\frac{1}{2}g^{2t}+\frac{1}{2}Ng^{1s}+Ng^{1t} \ ,\\
\end{split}
\end{equation}
explicitly demonstrating that double trace tensor structures are produced. This means that the double trace operators are needed for renormalisability. They are generated by one loop corrections.

In contrast, the algebra of double trace tensor structures closes:
\begin{equation}\label{eq:2trdiamond}
\begin{split}
g^{2s}  \diamond g^{2s} &= (M(N^2-1) +8)g^{2s}  
\\
g^{2t} \diamond g^{2t} &= (2M+2N^2+6)g^{2t} + 12 g^{2s} 
\\
g^{2s} \diamond g^{2t} &= 2(M+N^2)g^{2s}+6g^{2t} \ ,
\end{split}
\end{equation}
which means that it forms a sub-algebra.

Finally, the product of single trace and double trace tensor structures generates linear combinations of single trace and double trace structures,
\begin{equation}\label{eq:12trdiamond}
\begin{split}
g^{1s} \diamond g^{2s} &= (M+1)Ng^{2s} + 6g^{1s}
\\
g^{1s} \diamond g^{2t} &= 2(M+3)g^{1s} +2Ng^{2s} + Ng^{2t} + 8g^{1t} \\
g^{1t} \diamond g^{2s} &=  Ng^{2s} + 6g^{1t}  
\\
g^{1t} \diamond g^{2t} &= 2(M+1)g^{1t} + Ng^{2t} + 4g^{1s} .
\\
\end{split}
\end{equation}
Interestingly, these mixed products simplify considerably in the the large $N$ limit, where only double trace operators survive. Next, we describe this limit.

The standard large $N$ limit amounts to taking $N$ large while keeping rescaled 't Hooft couplings fixed, in our case $\lambda_{1S}=N\lambda_{1s}$, $\lambda_{1T}=N\lambda_{1t}$ and $\lambda_{2S}=N^2\lambda_{2s}$, $\lambda_{2T}=N^2\lambda_{2t}$. We have thus adopted the convention to use capital $S$ and $T$ symbols in place of $s$ and $t$ whenever they describe a 't Hooft coupling or a corresponding  large $N$ rescaled operator, tensor or polynomial. Hence, the rescalings 
\begin{equation} \label{eq_sum1}
\begin{split}
g^{1S}&= \frac{1}{N} {g}^{1s} \\
g^{1T} &= \frac{1}{N} {g}^{1t} \\
g^{2S} &= \frac{1}{N^2} g^{2s}\\
g^{2T}&= \frac{1}{N^2} g^{2t},
\end{split}
\end{equation}
preserve the decomposition \eqref{eq:decomposition}, making the large $N$ choice of variables a simple change of base. We then find the large $N$ algebra directly by studying the algebra of the rescaled operators, minding the transformation properties \eqref{eq:transcoeff} and dropping subleading terms:

\begin{equation}\label{eq:1TRdiamond}
\begin{split}
g^{1S} \diamond g^{1S} &=  \frac{1}{2}(M+3)g^{2S} + \frac{1}{2}g^{2T} + \frac{1}{2}(M+3)g^{1S} \\
g^{1T} \diamond g^{1T} & = \frac{1}{8}(M+2)g^{2T} + \frac{1}{2}g^{1S}\\
g^{1S} \diamond g^{1T} &=\frac{1}{2}g^{2S} + \frac{1}{2}g^{2T} + \frac{1}{2}g^{1S}+g^{1T} \\
\end{split}
\end{equation}
\begin{equation}\label{eq:2TRdiamond}
\begin{split}
g^{2S}  \diamond g^{2S} = &Mg^{2S} \\
g^{2T} \diamond g^{2T} =
&  2g^{2T}\\
g^{2S} \diamond g^{2T} =
& 2g^{2S} \\
\end{split}
\end{equation}
\begin{equation}\label{eq:12TRdiamond}
\begin{split}
g^{1S} \diamond g^{2S} =
&(M+1)g^{2S} \\
g^{1S} \diamond g^{2T} =
& 2g^{2S}+g^{2T} \\
g^{1T} \diamond g^{2S} =
&g^{2S} \\
g^{1T} \diamond g^{2T} =
&g^{2T}\\
\end{split}
\end{equation}

The algebra clearly simplifies in the large $N$ limit. 
As promised above, the mixed single trace-double trace products are all double trace tensor structures. From the previous result that double trace-double trace products yield double trace structures, we conclude that \emph{all} products of a double trace structure with anything else yields a double trace structure in the large $N$ limit. In algebra terminology, this amounts to the statement that the double trace subalgebra is in fact an algebra \emph{ideal}. 
Then, one may consistently mod out the double trace operators and construct a quotient algebra, which essentially is the algebra of the single trace operators, ignoring any double trace terms. Physically, the beta functions, of the single trace operators are independent of the double trace couplings, and the single trace RG equations form a closed dynamical system. See \eqref{eq_bd22} below. The quotient algebra controls the RG equations of $\lambda_{1S}$ and $\lambda_{1T}$ independently of $\lambda_{2S}$ and $\lambda_{2T}$.
As noted previously, the double trace sector of a large $N$ gauge theory is still significant \cite{Adams:2001jb,Dymarsky:2005nc,Dymarsky:2005uh,Pomoni:2009joh} since a running double trace coupling may prevent conformal fixed points or may lead the four point coupling to a region where the potential is unstable. There is also a smaller ideal spanned by $g^{2S}$ inside the double trace ideal. We then expect its quotient algebra to generate an RG flow independent of $\lambda_{2S}$, as can be confirmed by inspecting \eqref{eq_bd22} below.

Adding $g^{1S}$ to the double trace ideal still generates a sub-algebra, but not an ideal, due to the $g^{1T}$ term in $g^{1S} \diamond g^{1T}$. This means that a model with vanishing $\lambda_{1T}$ is consistent with the RG flow, and that there is no quotient algebra controlling the RG evolution of $\lambda_{1T}$ independently of $\lambda_{2S}, \lambda_{2T}$ and $\lambda_{1S}$. Furthermore, the tensor $g^{1T}$ does not form a sub-algebra by itself, much less an ideal, making the RG equations of $\lambda_{2S}, \lambda_{2T}$ and $\lambda_{1S}$ dependent on $\lambda_{1T}$.

We conclude this section by some other algebra related observations, without spelling out their physical implications here. Non-associative algebras with various properties, some mentioned below, are studied in their own right \cite{MR1375235}. In particular, commutative non-associative algebras like $\mathcal{A}$ have been used in the mathematics literature (starting with \cite{MR0132743}) for classifying systems of ordinary differential equations. Algebraic properties are directly related to the properties of the dynamical system defined by the algebra, which in our case would be the RG equations.

Commutative and non-associate algebras are not prominent in the theoretical physics toolkit, but Jordan algebras, with a $\circ$ product satisfying the Jordan property
$$
a^2\circ(a\circ b) = a \circ (a^2\circ b),
$$
where $a^2 \equiv a\circ a$, have been studied now and again. Another related property is power associativity, which means that $a^m$ is unambiguously defined and $a^m \circ a^n =a^{m+n}$ holds for all positive integers $m$ and $n$. 


The large $N$ double trace ideal is power associative, and has a unit element\footnote{Note that this element is not a unit element in the full algebra.} $\mathbb{I} = g^{2T}/2$. These facts imply that it is actually a Jordan algebra, and it is simple enough that we can give a complete characterisation. The element $\mathbb{E}_1 \equiv g^{2S}/M$ is idempotent, ie  $\mathbb{E}_1\diamond \mathbb{E}_1 = \mathbb{E}_1$. We also find that $\mathbb{E}_2 \equiv \mathbb{I} -\mathbb{E}_1$ is idempotent. These elements are then complementary projection operators, which means that this two-dimensional Jordan algebra decomposes into the direct sum $\mathbb{R} \oplus \mathbb{R}$.

We have found the full algebra $\mathcal{A}_{SU(N)}$ corresponding to symmetric $SU(N)$ adjoint scalar gauge theory to neither satisfy the Jordan property nor power associativity. It only satisfies the weaker \emph{flexible} identity
$$
a\diamond (b \diamond a) = (a \diamond b) \diamond a
$$
for arbitrary elements $a$ and $b$.

\section{Adjoint multi-scalar gauge theory beta functions}\label{sect:Adjointmsgbeta}

We can now find the complete beta function of the scalar couplings, and write it down in the large $N$ limit from our expressions for the algebra. We then discuss the relation to a simple bosonic D brane model, and then obtain its beta function both from the algebra and by a direct calculation. The D brane model is found to require completion to the full multi-adjoint gauge theory. All results agree with each other and with calculations in the Mathematica package RGBeta.

\subsection{Complete one loop beta functions from the algebra}

As seen in section \ref{sect:quartic} the algebra $\mathcal{A}$ directly reflects the one-loop beta functions \eqref{eq_bk}, which can be read off from its structure constants, which are extracted from the $\diamond$ products \eqref{eq:1trdiamond}, \eqref{eq:2trdiamond} and \eqref{eq:12trdiamond} or \eqref{eq:1TRdiamond}, \eqref{eq:2TRdiamond} and \eqref{eq:12TRdiamond}. We find the large $N$ beta functions in terms of large $N$ scalar 't Hooft couplings:
\begin{equation} \label{eq_bd22}
\begin{split}
\beta_{1S} &= \frac{1}{16\pi^2} \bigg( \frac{M+3}{2} \lambda_{1S}\lambda_{1S} + \frac{1}{2}\lambda_{1T}\lambda_{1T} + \lambda_{1T}\lambda_{1S} -12Ng^2\lambda_{1S} +6 N^2g^4 \bigg) \\
\beta_{1T} &=  \frac{1}{16\pi^2} \bigg( 2\lambda_{1T}\lambda_{1S} -12Ng^2\lambda_{1T} \bigg) \\
\beta_{2S}
 &=\frac{1}{16\pi^2} \bigg(M \lambda_{2S}\lambda_{2S}  + \frac{(M+3)}{2}\lambda_{1S}\lambda_{1S} +4\lambda_{2S}\lambda_{2T} + 2\lambda_{2S}\lambda_{1T} +2(M+1)\lambda_{2S}\lambda_{1S}+4\lambda_{2T}\lambda_{1S} \\&+\lambda_{1T}\lambda_{1S}  
-12Ng^2\lambda_{2S} + 6N^2g^4  \bigg) \\
\beta_{2T} &= \frac{1}{16\pi^2} \bigg( 2\lambda_{2T}\lambda_{2T} + \frac{1}{2}\lambda_{1S}\lambda_{1S} + \frac{M+2}{8}\lambda_{1T}\lambda_{1T} +2\lambda_{2T}\lambda_{1T} + 2\lambda_{2T}\lambda_{1S}+\lambda_{1T}\lambda_{1S} \\
&-12Ng^2\lambda_{2T} +6N^2g^4\bigg). \\
\end{split}
\end{equation}
Note that the $\lambda_{1T}$ coupling can self-consistently be set to zero. The reason is twofold. First, the gauge boson loops do not induce running coupling with a $g^{1T}$ structure. Second, the tensors corresponding to the other couplings form a closed algebra, and will not generate a $g^{1T}$ term at one loop. We indeed find below that all the most interesting features of the RG flow occur at $\lambda_{1T}=0$.

\subsection{A D-brane model}

The adjoint multi-scalar gauge theories we study are related to string theory D-branes \cite{DaiLeighPolchinski1989}, which are defects where open strings can end. The low energy effective theory for open strings on $N$ D-branes is $U(N)$ Yang-Mills theory with $M$ adjoint scalars representing fluctuations in dimensions transverse to the defect. There are also typically massless fermions required by supersymmetry. In many cases, but not all, with insufficient supersymmetry, there are tachyons in the string theory, which render it unstable. Purely bosonic string theory suffers from this problem, which is why we call the corresponding gauge theories D-brane models rather than limits of the theory on the D-brane. Although it is possible that the theory becomes meta-stable in the field theory limit we do not make such claims, but use the formal limit without tachyons as an interesting example.

The classical scale invariant four point coupling for the scalars in these models is completely analogous to the Yang-Mills four point coupling, and given by dimensional reduction of higher-dimensional Yang-Mills theory. In the maximally supersymmetric case, the coupling receives no quantum corrections and scale invariance is preserved. The potential has flat directions but is otherwise positive \cite{Witten1996}. In more general cases, these D brane models are still renormalisable in four dimensions, but will be corrected by loops and a scale will typically be introduced through dimensional transmutation. Generically, the potential lacks flat directions and is stable or unstable corresponding to favouring $N$ branes staying together or moving apart, respectively. Since the classical behaviour is marginally stable, we expect the first quantum corrections, which we study here, to be decisive.

The classical four-point coupling is modelled on the Yang-Mills four-point coupling, and reads 
\begin{equation}\label{eq_lambda}
\begin{split}
\lambda_{aAbBcCdD} 
&=-\frac{\lambda}{6}[\delta_{a c} \delta_{b d} f_{A B}^{E} f_{C D}^{E}+\delta_{b c} \delta_{a d} f_{B A}^{E} f_{C D}^{E}+\delta_{c a} \delta_{b d} f_{C B}^{E} f_{A D}^{E} \\
&+\delta_{d c} \delta_{b a} f_{D B}^{E} f_{C A}^{E}+\delta_{c b} \delta_{a d} f_{C A}^{E} f_{B D}^{E}+\delta_{d c} \delta_{a b} f_{D A}^{E} f_{C B}^{E}] \\
&= -\frac{\lambda}{3} \big[ (\Tr(T_AT_BT_CT_D)+\Tr(T_AT_DT_CT_B))(2\delta_{ac}\delta_{bd}-\delta_{ab}\delta_{cd}-\delta_{ad}\delta_{bc}) \\
&+(\Tr(T_AT_BT_DT_C)+\Tr(T_AT_CT_DT_B))(2\delta_{ad}\delta_{bc}-\delta_{ab}\delta_{cd}-\delta_{ac}\delta_{bd}) \\
&+(\Tr(T_AT_CT_BT_D)+\Tr(T_AT_DT_BT_C))(2\delta_{ab}\delta_{cd}-\delta_{ad}\delta_{bc}-\delta_{ac}\delta_{bd})  \big] \\
&= \frac{\lambda}{3}(g^{1s}-2g^{1t})
\end{split}
\end{equation}
in terms of the Lie algebra structure coefficients $ f^A_{BC}$ from $\left[ T_B, T_C\right] = i f^A_{BC} T_A$.

\subsection{D-brane model beta functions from the general theory}
Focusing on leading terms for large $N$, the D-brane model's beta functions follow from the general beta function \eqref{eq_bd22} by specialising the four point coupling to \eqref{eq_lambda}. 
Alternatively, the standard expressions \eqref{eq:beta} and \eqref{eq_LLA} may be used directly, or we use the general beta functions \eqref{eq_antz} from the components \eqref{eq_bk} and the products \eqref{eq:1trdiamond} with $\lambda_{i}$ from the D-brane model interaction \eqref{eq_lambda}:
\begin{equation} \label{eq_bi}
\begin{split}
\mathbf{\beta}^{I}_{\bar{A}\bar{B}\bar{C}\bar{D}} 
&=  g^{1s} \bigg( \frac{N(M+3)\lambda^2}{18} -4Ng^2\lambda + 6g^4N \bigg) + g^{1t}\bigg( \frac{-4N\lambda^2}{9} +8Ng^2\lambda \bigg) \\ &+ (g^{2s}+g^{2t}) \bigg( \frac{(M-1)\lambda^2}{18} +6g^4\bigg),
\end{split}
\end{equation}
which to this order yields
\begin{equation} \label{eq_bdff}
\begin{split}
\beta_{1s} &= \frac{1}{16\pi^2}  \bigg( \frac{N(M+3)\lambda^2}{18} -4Ng^2\lambda + 6g^4N \bigg) \\
\beta_{1t} &= \frac{1}{16\pi^2} \bigg( \frac{-4N\lambda^2}{9} +8Ng^2\lambda \bigg) \\
\beta_{2s} &=  \frac{1}{16\pi^2}  \bigg(\frac{(M-1)\lambda^2}{18} +6g^4\bigg) \\
\beta_{2t} &= \frac{1}{16\pi^2} \bigg( \frac{(M-1)\lambda^2}{18} +6g^4\bigg). \\
\end{split}
\end{equation}
Double trace tensor structures are generated, and the two single trace couplings run at different speeds. This means that the bosonic D brane model is not consistent with the RG flow and has to be extended to four coupling constants. In the end, a specific RG trajectory could describe a gauge theory descending from string theory, but which particular trajectory is not a pure field theory question. Furthermore, the general one loop RG flows we find below in section \ref{sect:RGFlow} do not support the idea that a tree level coupling \eqref{eq_lambda} can describe the model over a substantial range of scales.

We have also obtained the same beta functions \eqref{eq_bdff} directly, using standard birdtrack methods \cite{Cvitanovic:1976am,Cvitanovic:2008zz}, thereby partially validating our algebra approach. The result also agrees with a calculation in the Mathematica package RGBeta \cite{Thomsen:2021ncy}\footnote{We thank Fedor Popov for alerting us to a discrepancy between an RGBeta calculation and an earlier edition of our calculation. The results and methods are now thoroughly cross-validated.}.

\section{One loop RG flow in adjoint multi-scalar gauge theory}\label{sect:RGFlow}

We can now discuss the physics of the RG flow and how it follows from the algebra of invariants. We consider the one loop RG flow equations and the partial decoupling of the gauge coupling by focusing on ratios of scalar and gauge couplings. Then we describe fixed points of these ratios, which are fixed flows of the original scalar couplings. The full structure of real fixed flows is surveyed for positive $M$ in the large $N$ limit. A hierarchical structure of the RG equations in this limit simplifies the investigation. In particular, the RG flow of the single trace couplings is independent of double trace equations. We also briefly describe complex fixed points and flows.

\subsection{The algebra and the RG flow}

The beta functions above describe the RG flow, but it should be noted that  the gauge coupling $g^2$ and the corresponding 't Hooft coupling $N g^2$ are usually not constant but running. The running gauge coupling is taken into account simply by a new choice of relative couplings $\mu_i$ with RG equations that form an autonomous system of first order equations. The idea to focus on the fixed points of the relative couplings, called \emph{fixed flows} was advocated a while ago \cite{Giudice:2014tma} and has a long history \cite{Gross:1973ju,Cheng:1973nv}. For fixed flows scalar couplings are locked relative to each other and the gauge coupling. The complete system of one loop RG equations includes the gauge coupling beta function, which we write in terms of $\beta_{g^2}$ or $\beta_{Ng^2}$:
\begin{equation}
\beta_{g}=-\frac{g^3}{16\pi^2}\frac{22-M}{6} C_2(G)
\end{equation}
\begin{equation}\label{eq:g2beta}
\beta_{g^2}=-2g\frac{g^3}{16\pi^2}\frac{22-M}{6} C_2(G)
\end{equation}
\begin{equation}\label{eq:g2beta}
\beta_{Ng^2}=-\frac{22-M}{48\pi^2} (Ng^2)^2,
\end{equation}
where we used $C_2(G)=N$.
As sketched above, we disentangle the dependence of the RG equations on the gauge coupling for non-critical $M\neq22$ by considering the relative (large $N$) scalar couplings $\mu_I \equiv \frac{\lambda_I}{Ng^2}
$ 
which yields their beta functions
\begin{equation} \label{eq:relbeta}
\begin{split}
\beta_{1S} (\mu)&= \frac{Ng^2}{16\pi^2} \bigg( \frac{M+3}{2} \mu_{1S}\mu_{1S} + \frac{1}{2}\mu_{1T}\mu_{1T} + \mu_{1T}\mu_{1S} -(\frac{M-22}{3}+12)\mu_{1S} +6  \bigg) \\
\beta_{1T}(\mu) &=  \frac{Ng^2}{16\pi^2} \bigg( 2\mu_{1T}\mu_{1S} -(\frac{M-22}{3}+12)\mu_{1T} \bigg) \\
\beta_{2S} (\mu)
 &=\frac{Ng^2}{16\pi^2} \bigg(M\mu_{2S}\mu_{2S}  + \frac{(M+3)}{2}\mu_{1S}\mu_{1S} +4\mu_{2S}\mu_{2T} + 2\mu_{2S}\mu_{1T} +2(M+1)\mu_{2S}\mu_{1S}+4\mu_{2T}\mu_{1S} \\&+\mu_{1T}\mu_{1S}  
-(\frac{M-22}{3}+12)\mu_{2S} + 6  \bigg) \\
\beta_{2T}(\mu) &= \frac{Ng^2}{16\pi^2} \bigg( 2\mu_{2T}\mu_{2T} + \frac{1}{2}\mu_{1S}\mu_{1S} + \frac{M+2}{8}\mu_{1T}\mu_{1T} +2\mu_{2T}\mu_{1T} + 2\mu_{2T}\mu_{1S}+\mu_{1T}\mu_{1S} \\
&-(\frac{M-22}{3}+12)\mu_{2T} +6\bigg). \\
\end{split}
\end{equation}
Together with the gauge coupling beta function \eqref{eq:g2beta} these beta functions give a complete set of RG equations. Note that the one loop RG equations have the form $\frac{d\mu}{dt}=\alpha f(\mu), \frac{d\alpha}{dt}=-b \alpha^2$, which allows them to be completely separated by a new choice of RG time given by $d\tau =\alpha dt$ into $\frac{d\mu}{d\tau}=f(\mu)$ and $\frac{d\alpha}{d\tau}=-b \alpha$. This means that the RG flow phase portrait in $\mu_i$ space is entirely independent of the flow of the gauge coupling $\alpha$ and forms a separate dynamical system. We investigate this system by regarding the gauge coupling as constant.

\subsection{Real or complex fixed flows}
We now search for fixed flows in the large $N$ limit by setting the beta functions of relative couplings to zero, $\beta_i(\mu)=0$. At one loop these equations are four second order polynomial equations in four variables which simplify for large $N$ because the the single trace equations based on the large $N$ beta functions \eqref{eq:relbeta} only involve the single trace couplings.
The double trace flows then also simplify considerably for fixed single trace flows. This is because the flow of the double trace couplings is parametrised by the single trace couplings, which are then constant. Our procedure to analyse the RG flow makes use of this hierarchical structure. The actual flow pattern is then determined by number $M$ of adjoint scalars.

In practice, the phase portraits of the single trace flow are of primary importance and can be analysed directly, then yielding different phase portraits of double trace flows for each single trace fixed flow. We focus especially on the completely IR or UV stable flows.

The single trace fixed flow equations are also hierarchically arranged and easy to solve, first solving for $\mu_{1T}$ in terms of $\mu_{1S}$, and then solving the resulting second order equation. We find that there are two pairs of single trace fixed flow (FF) roots, $\mu^{FF1\pm}$ and $\mu^{FF2\pm}$:
\begin{equation}\label{eq:1trFF}
\begin{array}{clr}
\mu_{1S}^{FF1\pm}&=\frac{M+14\pm \sqrt{M^2-80 M-128}}{3 (M+3)},& \mu_{1T}^{FF1\pm}= 0    \\
 && \\
\mu_{1S}^{FF2\pm}&=\frac{M+14}{6},& \mu_{1T}^{FF2\pm}=-\frac{M+14}{6}\pm \frac{\sqrt{-M^3-26 M^2-140 M-40}}{6} . 
\end{array}
\end{equation}
None of the second pairs of fixed flow roots are real for the physical $M>0$. The first pair of fixed flow roots is more interesting: There are two single trace fixed flows, given by the first pair of roots $\mu^{FF1\pm}$ for integer $M\ge 82$. An important conclusion is that there are \emph{no real one-loop fixed points} for this class of theories. The fixed flow equations above are fixed point equations only for $M=22$, which is the conditions for the gauge coupling to not flow at one loop order. That $M=22$ is outside the range which allows real fixed flows means that there are no real fixed points to this order. 

Here, we pause the general search of real fixed flows to temporarily consider complex roots. Even if complex coupling constants in general break unitarity complex roots of the fixed point equations are interesting in their own right. They define complex conformal field theories \cite{Gorbenko:2018ncu}, and similarly, the fixed flows at complex values of the roots above are complex fixed flows. Such flows give characteristic flow patterns for real couplings in the vicinity of the complex roots \cite{Gorbenko:2018ncu,Faedo:2021ksi}, as we will see an example of in figure \ref{fig:2TRNoFF82}. Allowing for complex roots, all the four single trace fixed flows \eqref{eq:1trFF} make sense, and each one of them gives rise to a second order equation for $\mu_{2T}$. Finally, each of these roots yields a second order equation for $\mu_{2S}$. Altogether, there are generically 16 complex roots arranged hierarchically, and for each set of complex roots, the complex conjugate couplings also solve the equations. Specifically, the case $M=22$ yields complex valued fixed points, and in contrast to related examples engineered from $\mathcal{N}=4$ supersymmetric Yang-Mills theory, for which only double trace couplings are complex, one of the single trace couplings are complex in this purely bosonic case.  

\begin{figure}
\centering   
          \begin{subfigure}[a]{0.45\textwidth}
         \centering
         \includegraphics[width=\textwidth]{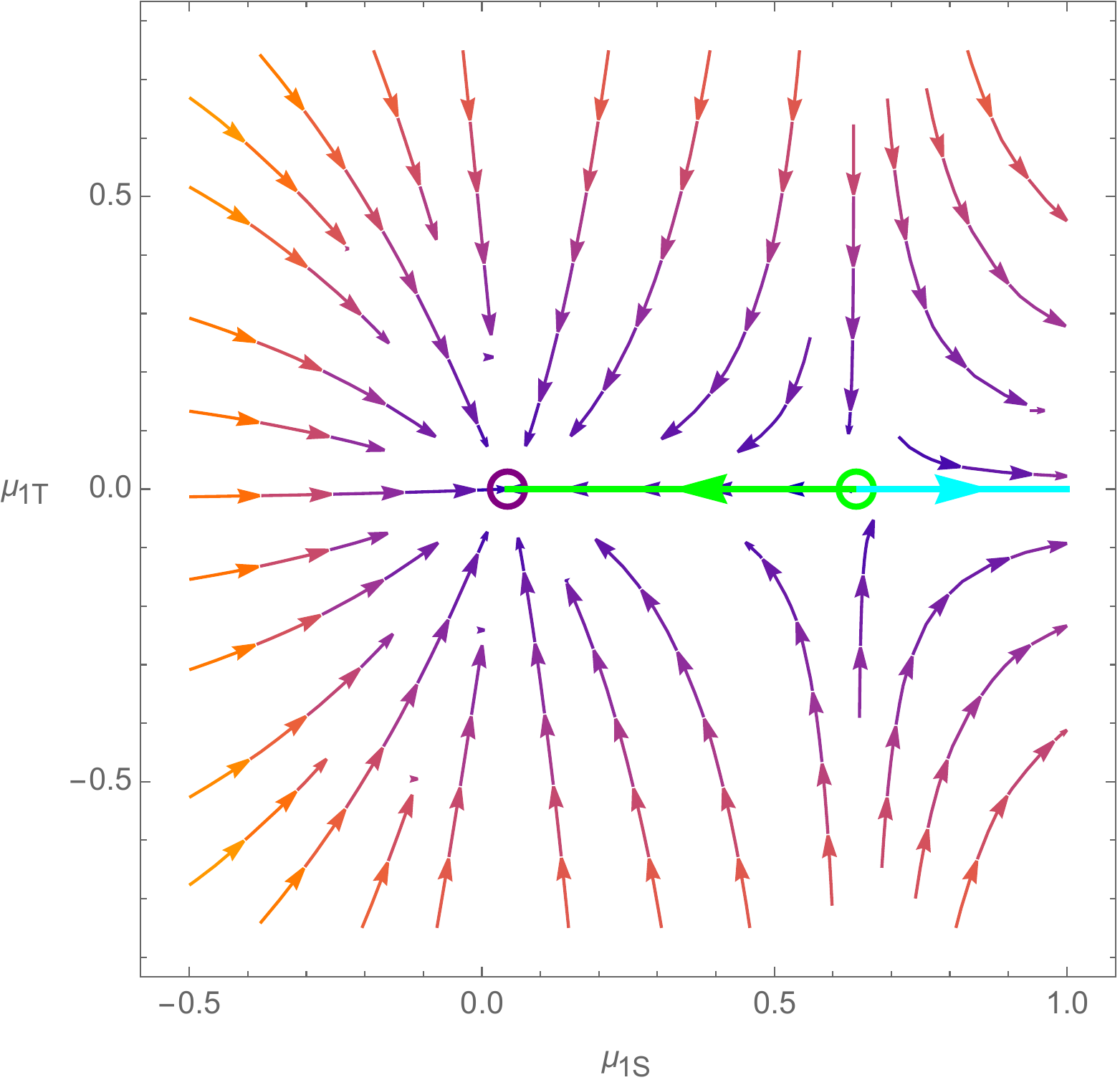}
         \caption{Single trace flow. Fixed flows determine parameters for double trace flows.}
         \label{fig:1TR425}
     \end{subfigure}
		\hfill
     \begin{subfigure}[a]{0.45\textwidth}
         \centering
         \includegraphics[width=\textwidth]{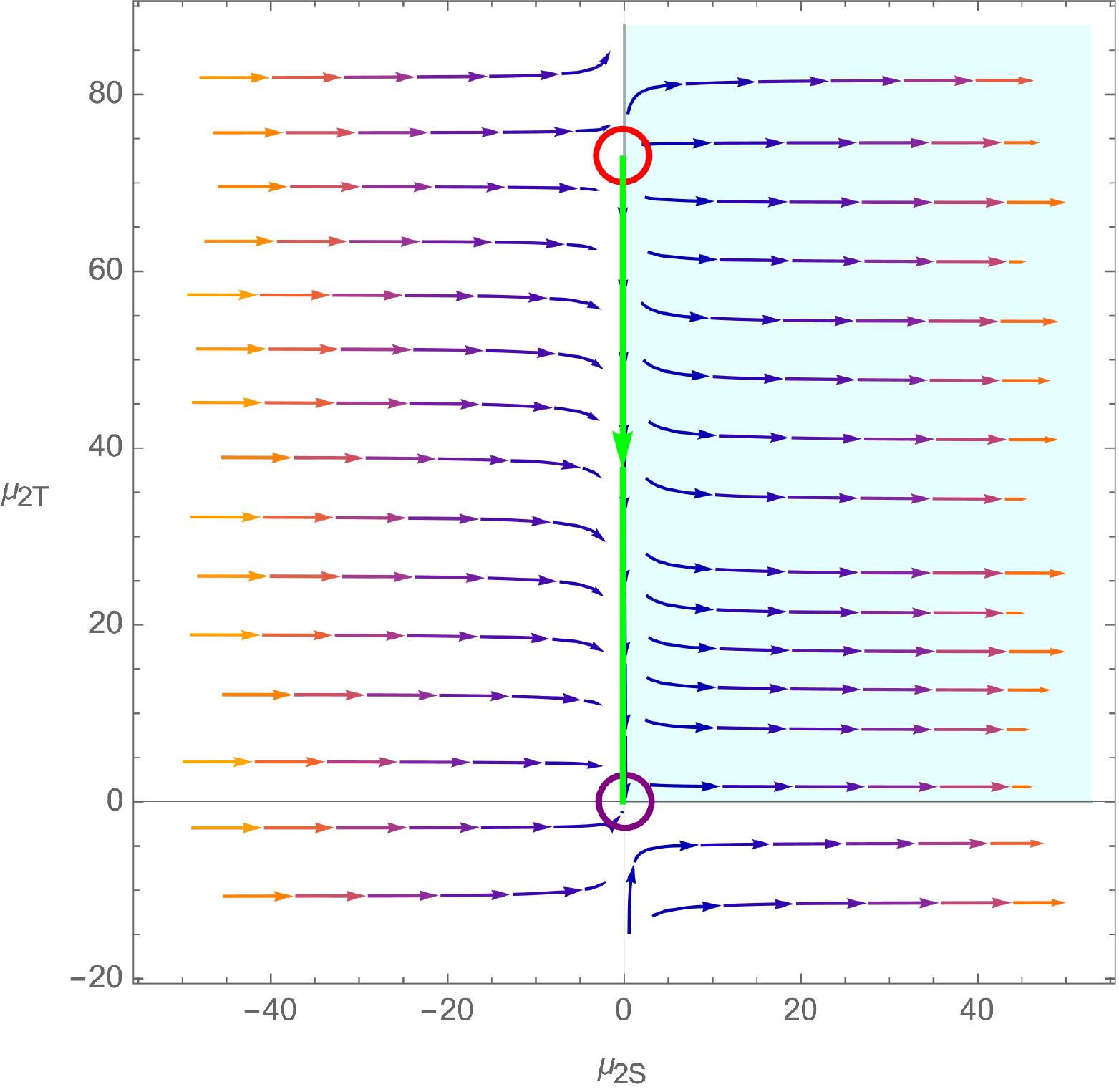}
         \caption{Double trace flow, coarse grained view, Details are resolved in figure \ref{fig:2TRcloseups}.}
         \label{fig:2TRoverview}
     \end{subfigure}
        \caption{RG flows for $M=425$. UV stable fixed flows are marked with purple circles, IR stable fixed flows with red circles and fixed flows of mixed stability with green circles. They are connected by (green) separatrix flows. Cyan flows and regions mark other models for which the one loop approximation also makes sense in the IR.}
        \label{fig:PhasePortraits}
\end{figure}

Returning to the search for real fixed flows, the RG flows of the single trace couplings change qualitatively between $M=81$ and $M=82$, by the appearance of two real roots. In the lower range of $M$ there are only complex fixed flows, while in the upper range there are four special flows: a UV stable fixed flow, an unstable fixed flow,  a separatrix flow between the two fixed flows, and one between the unstable fixed flow and infinity. Figure \ref{fig:1TR425} illustrates a typical phase portrait for single trace flows with $M\ge82$. Note that all the theories described by these phase portraits are IR free in terms of the gauge coupling. Despite the IR weakness of the gauge couplings, the scalar single trace couplings generally grow in the IR, with the exception of these four special flows. The single trace couplings and the gauge coupling are asymptotically IR free for these four flows, such that the one loop approximation is expected to hold.

The final verdict on the consistency of the one loop approximation in the IR is settled by a study of the double trace phase portraits. First, the single trace fixed flow roots determine quadratic fixed flow equations for the double trace couplings. There are no completely real double trace roots for $M \le 405$. The qualitative behaviour is illustrated in figure \ref{fig:2TRNoFF82}. 
For $M= 406, \dots, 427$ and the UV stable single trace root $\mu^{FF1-}$,  there are  four fixed flows at real couplings, while for $M\ge 428$ there are two groups of four real fixed flow solutions, one for each of the two real single trace fixed flows. 
\begin{figure}
\centering
     \begin{subfigure}[t]{0.3\textwidth}
         \centering
         \includegraphics[width=\textwidth]{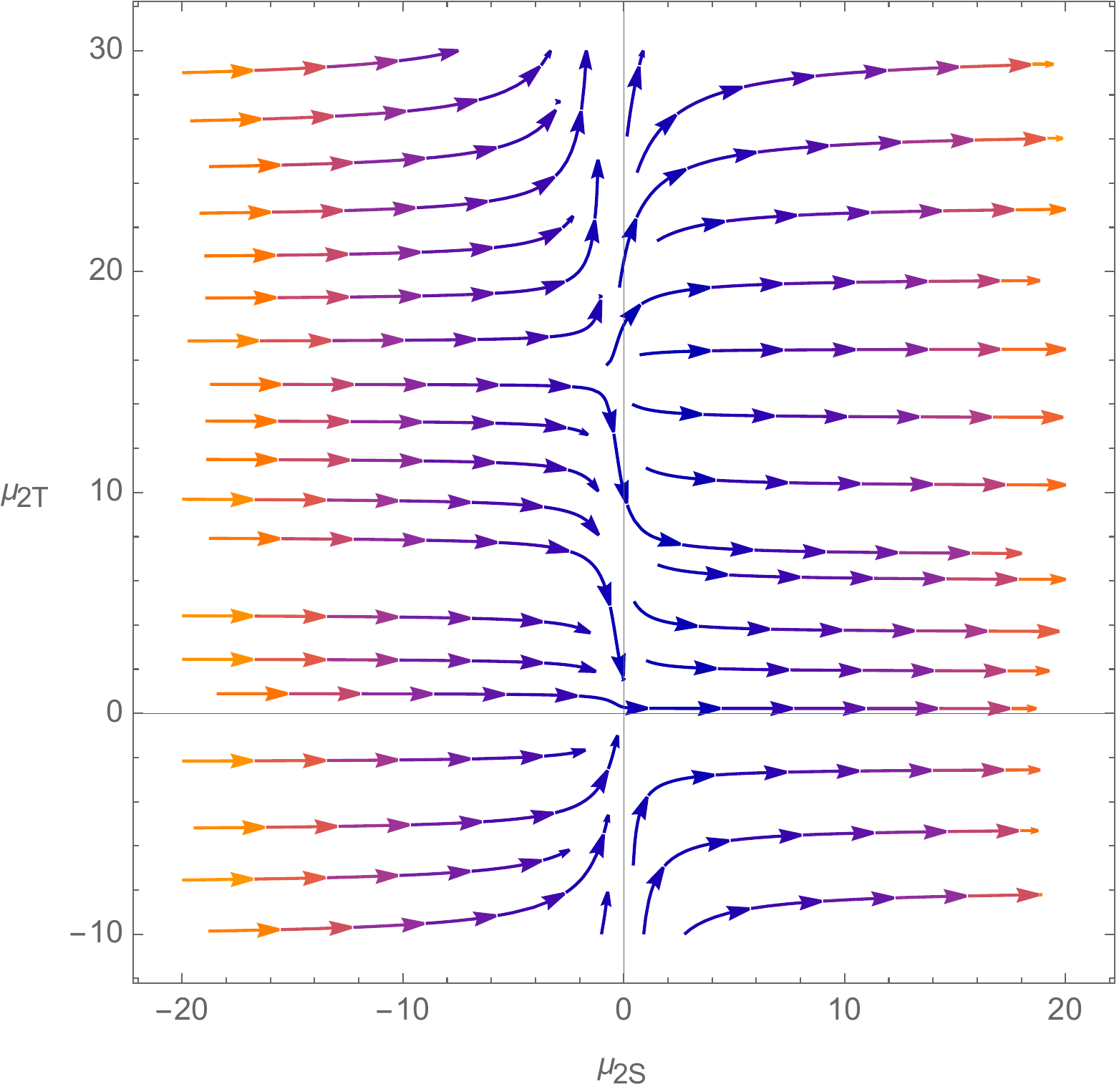}
         \caption{Overview.}
         \label{fig:2TRNoFFOverview}
     \end{subfigure}
\hfill
\begin{subfigure}[t]{0.3\textwidth}
\centering
\includegraphics[width=\textwidth]{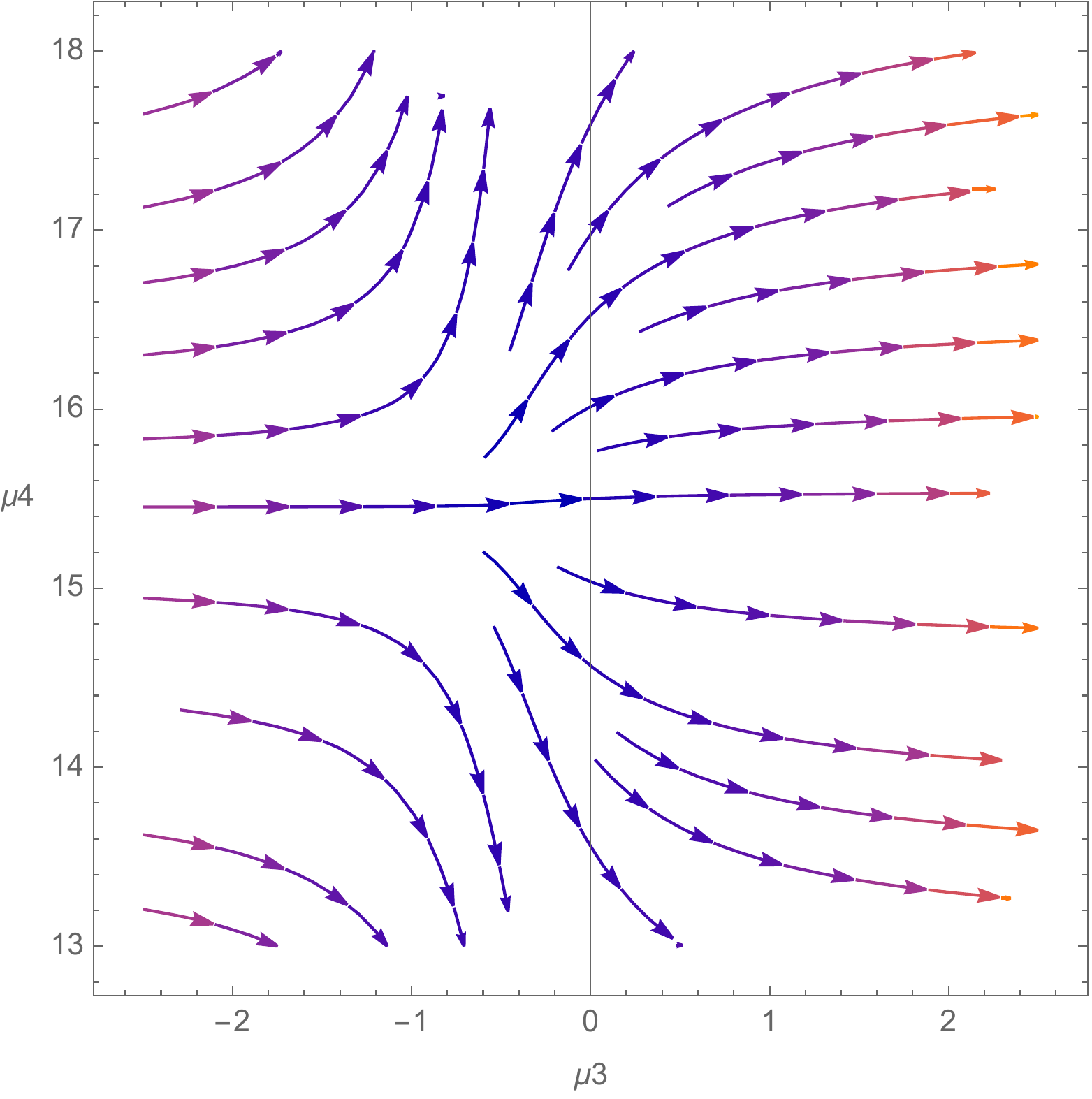}
\caption{Upper flow feature.}
\label{fig:Upper2TRNoFF}
\end{subfigure}
\hfill		
     \begin{subfigure}[t]{0.3\textwidth}
         \centering
         \includegraphics[width=\textwidth]{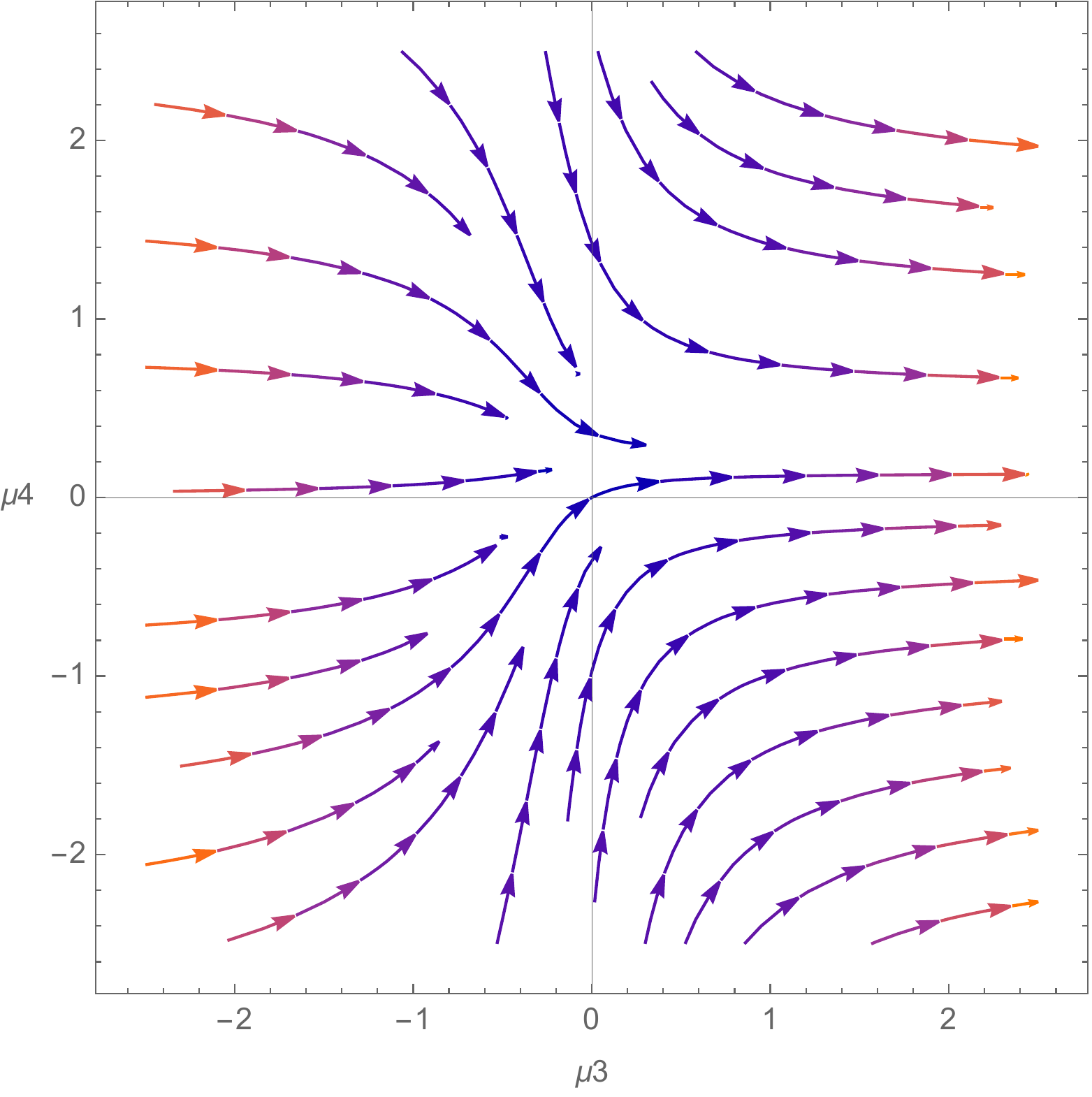}
         \caption{Lower flow feature.}
         \label{fig:Lower2TRNoFF}
     \end{subfigure}
        \caption{Double trace flows without fixed flows for $M=82$. Compare the coarse grained view of figure \ref{fig:2TRoverview} and closeups of the flows around its fixed flows given below in figure \ref{fig:2TRcloseups}. The features above represent the presence of two complex conjugate fixed flows, which coalesce between $M=405$ and $M=406$ and separate again for large $M$ into the pairs of fixed points of figure \ref{fig:2TRcloseups}.}
        \label{fig:2TRNoFF82}
\end{figure}

\begin{figure}
\centering   
          \begin{subfigure}[a]{0.45\textwidth}
         \centering
         \includegraphics[width=\textwidth]{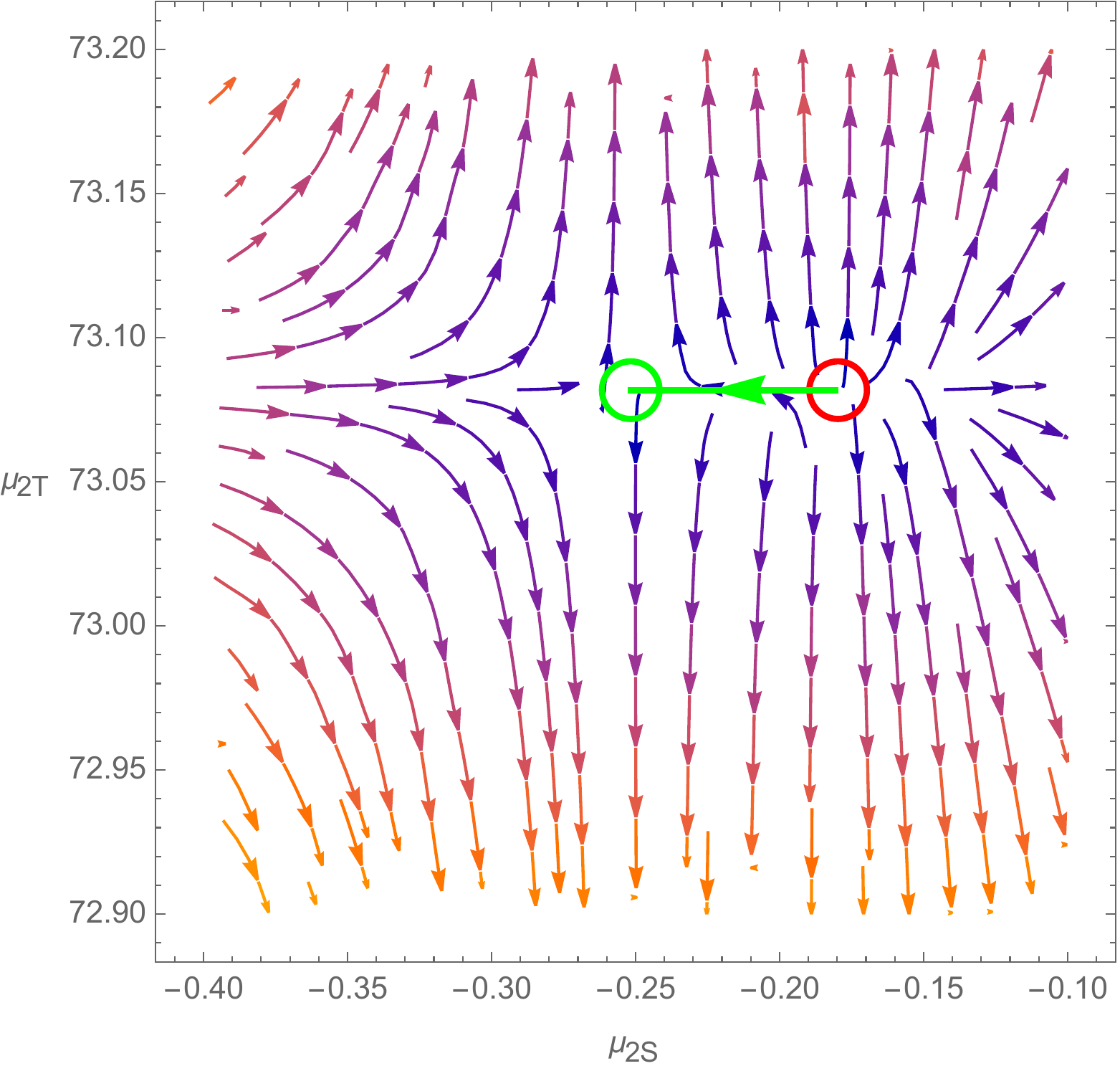}
         \caption{Upper fixed flow details.}
         \label{fig:Upper2TR}
     \end{subfigure}
		\hfill
     \begin{subfigure}[a]{0.45\textwidth}
         \centering
         \includegraphics[width=\textwidth]{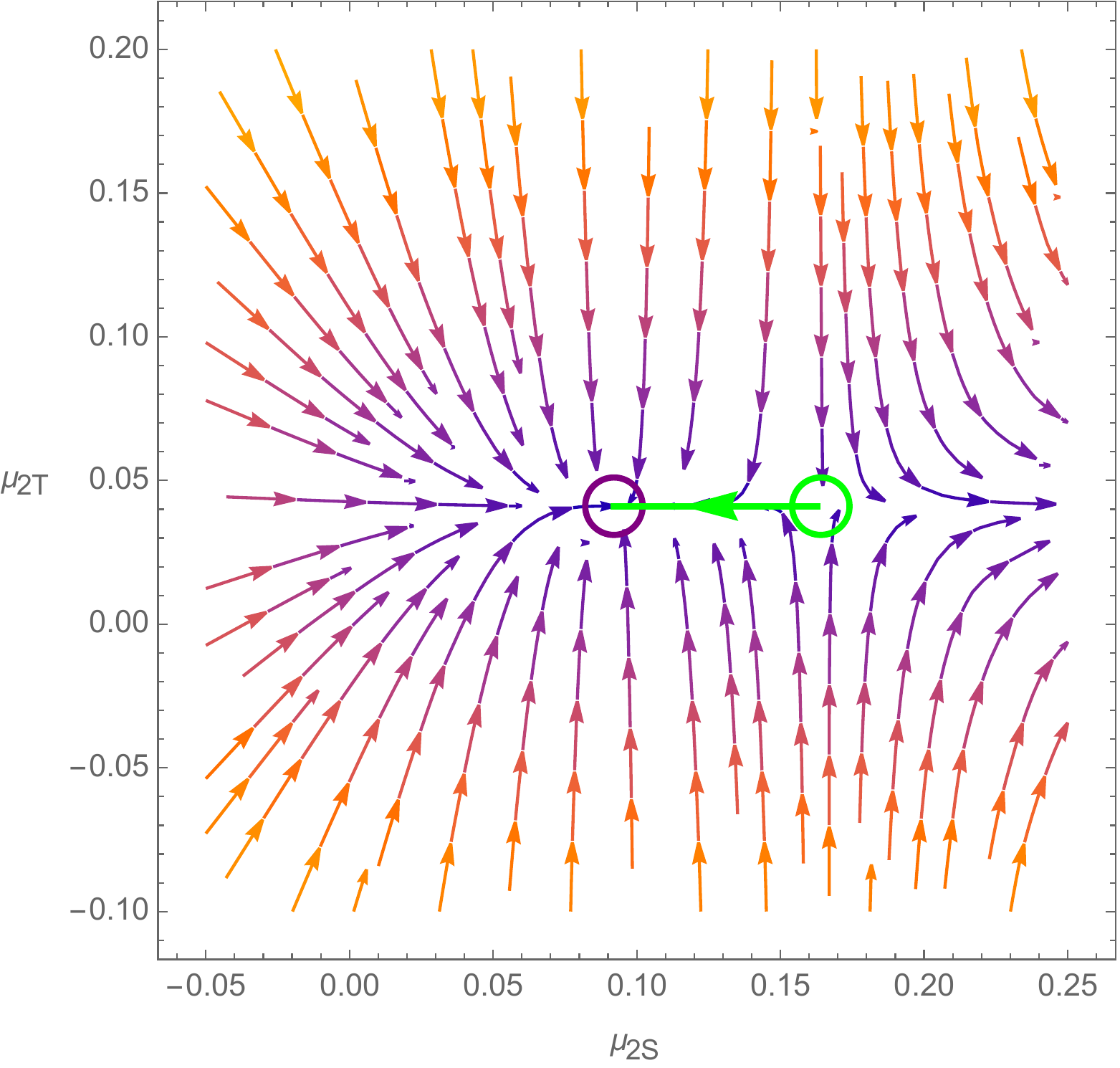}
         \caption{Lower fixed flow details.}
         \label{fig:Lower2TR}
     \end{subfigure}
        \caption{Close-ups of double trace flows for $M=425$ with fixed flows marked by circles. Two almost vertical separatrices connecting the upper and lower fixed flows have not been drawn. Together, the four fixed flows and the four separatices form a parallelogram through which no flows can pass. In the coarse grained view of figure \ref{fig:2TRoverview}, it appears as a separatrix.}
        \label{fig:2TRcloseups}
\end{figure}

The double trace phase portraits are all qualitatively captured by figures \ref{fig:2TRoverview} and \ref{fig:2TRcloseups} where the four fixed flows form tall, narrow parallelograms. In figure \ref{fig:2TRoverview} the parallelogram is not resolved and the general flow inside it is well represented by a separatrix. Except for the parallelogram/separatrix preventing any flows to pass through, the phase portraits \ref{fig:2TRNoFF82} for $M\le 405$ and \ref{fig:2TRoverview} for $M\ge 406$ are qualitatively similar. The obstacle to the flow for $M\ge 406$ however permits a whole class of trajectories to have asymptotically IR free limits, characterised by the IR stable double trace fixed flow in the phase portrait \ref{fig:2TRoverview}. The region covered by these flows has been coloured cyan.

For $406 \le M \le 427$, the only IR free double trace flows correspond to the UV stable single trace fixed flows. If the single trace flow is not exactly on this fixed point the IR limit will be strongly coupled, or end up on the separatrix connecting to the fixed flow of mixed stability in the IR. Since there are no double trace fixed flows associated with this single trace fixed flow in this range of $M$, its overall IR free status is undermined. Thus, for $406 \le M \le 427$ only the UV stable (IR unstable!) single trace fixed flow leads to IR free limits with the double trace flows illustrated by figure \ref{fig:2TRoverview}. In contrast, for $M \ge 428$, all the four specially marked fixed flows and separatrices in figure \ref{fig:1TR425} are compatible with such IR free limits.

We now summarise the big picture. Since we have only found fixed flows for $M \ge 406$, where the gauge coupling grows indefinitely in the UV, the one loop analysis yields no theory which is under control in the UV. Leaving short distance consistency aside, 
each trajectory which asymptotes to a fixed flow in the IR provides a model which is under perturbative control in the IR. For the fixed flows and bounded separatrices discussed above, the scalar sector is under complete control in the one-loop approximation, but the indefinitely growing gauge coupling ruins the one-loop approximation in the UV. Additional flows which are under control in the IR are coloured cyan in figure \ref{fig:PhasePortraits}.

For full IR consistency, the effective potential corresponding to the IR limit of the theory should be stable. We have not performed this analysis, but note that an unstable potential in the IR is likely to correspond to a Coleman-Weinberg symmetry breakdown \cite{Coleman:1973jx}. Many of the fixed flow couplings are actually positive and multiply positive semi-definite polynomials in $\Phi_a$, suggesting that at least some of the fixed flows represent stable potentials giving stable symmetric vacua. Such fixed flows then represent fully consistent symmetric theories in the IR. 

\section{Conclusion}\label{sect:conclusion}

\subsection{Discussion}

We have found the one-loop RG flow to be of limited validity. Only for very exceptional trajectories can the one-loop approximation be valid asymptotically in the infrared, and no trajectories are under control asymptotically in the ultraviolet. 

A fixed point of all couplings could have connected multiscalar adjoint gauge theory to a bosonic D3-brane model of potential use in a bosonic AdS/CFT correspondence through its large $N$ limit. At least our naive choice of gauge theory seems to fail. First, the initially promising coincidence of the critical dimension 26 for the bosonic string and the vanishing of the one-loop gauge coupling beta function in $D=4$ for $M=22=26-4$\footnote{On 3-branes in 26 dimensions live four-dimensional field theories with 22 scalar multiplets corresponding to fluctuations in 22 transverse dimensions.} does not survive the corrections from interactions: The one-loop beta functions of the scalar self-interactions have no zero for $M \le 405$. Second, one might have thought that the gauge theory D brane model in $M$ transverse dimensions could have had a chance to describe something similar to supersymmetric D branes in the large $M$ limit. We see no such signs. Allowing for running couplings, the one-loop flows of couplings seem to favour D brane interactions that are very different from the tree level interactions that govern supersymmetric D branes. There is no sign that the single trace flows derived from \eqref{eq_bd22} and displayed in figure \ref{fig:1TR425} retain any similarity to the tree level coupling over any reasonable range of scales. If these gauge theory models are related to D branes, the relation is different, or D brane physics is different from what we have assumed here. This may not come as a surprise, since we have neglected the tachyon instability that is present in a full string theory description of bosonic D branes, but the mismatch with the one loop RG evolution provides a another perspective on bosonic D branes.

Conformal invariance is ruined in our case, just as in the orbifolded gauge theories discussed in the introduction, unless we consider $M=22$, and accept complex couplings at the fixed points, ie non-unitarity complex CFTs. A new phenomenon is that not even single trace fixed point couplings are protected from being complex in our models. A basic question raised in \cite{Dymarsky:2005uh,Pomoni:2009joh} however remains. Conformal symmetry, or unitarity, is broken in all non-supersymmetric large $N$ conformal field theories with adjoint scalars that have been studied. Is a principle at work here, or have we just not found the right example yet?

Methodologically, we have introduced a non-associative algebra representation of the one loop renormalisation group. The results of this paper were computed in terms of the algebra but also checked both manually and with the RGBeta Mathematica package to standard calculations from general expressions \cite{Machacek:1984zw}.

\subsection{Outlook}

A more general validity of perturbative RG trajectories would require higher order corrections, but corrections could also reinforce the limitations of perturbation theory. It would be interesting to find out the fate of the renormalisation group phase portraits.

For the D3-brane interpretation of the adjoint scalar gauge theories, this study points to an impasse. It seems unlikely that higher corrections balance the one loop terms to yield a scalar self interaction that resembles the tree level interaction in a substantial part of the RG flow.  
A study of the effective potential rather than the four point coupling is however more relevant for the infrared flow, and could evade our negative conclusions. 
Another intriguing avenue would be to study the complex one loop fixed points for $M=22$ and regard them as a starting point for non-unitary bosonic D-branes and complex holography \cite{Faedo:2019nxw} in a complex version of AdS/CFT.

We have limited the present study to the RG evolution of the couplings, but a stability analysis of the potentials at fixed points is definitely warranted for all $M$. Even if the the RG trajectories are unstable in the infrared, vacuum stability can be investigated. Again, the effective potential is important. RG trajectories leading to instability of the naive vacuum should correspond to the instability of the effective potential to radiative symmetry breakdown in the Coleman-Weinberg mechanism \cite{Coleman:1973jx}, but it would be interesting to see this in detail. 

Finally, we are curious about the prospects of non-associative algebras illuminating the RG group in other systems, as well as to higher orders. Generalising our approach to other systems seems straightforward, since we merely rephrased well established equations in an algebraic language. The question is more if the algebraic formulation, or its application to dynamical systems leads to new insights. Perhaps, special quantum field theories instead illuminates mathematics. Generalising the non-associative algebras to higher orders seems much more challenging, but also potentially more rewarding.

\acknowledgments
We wish to thank Igor Klebanov, Fedor Popov, Francesco Sannino, Anders Eller Thomsen and Arkady Tseytlin for helpful correspondence. The work of B.S. is supported by the Swedish research council VR, contract DNR-2018-03803.

\bibliographystyle{JHEP}
\bibliography{Bibliography}

\end{document}